\documentclass{article}%
\usepackage{amsmath}
\usepackage{cite}
\usepackage{geometry}
\usepackage{amsfonts}
\usepackage{amssymb}
\usepackage{graphicx}%
\providecommand{\U}[1]{\protect\rule{.1in}{.1in}}
\pdfoutput=1

\begin{document}

\title{ }

\begin{center}
{\LARGE Analytic Computer Generated Hologram Design}

Gregg M. Gallatin

Applied Math Solutions, LLC

Newtown, CT

gregg@appliedmathsolutions.com

\bigskip
\end{center}

\section{Abstract}

Computer Generated Holograms (CGH), also known as Diffractive Null Correctors
(DNC) or Diffractive Null Lenses (DNL) are used to convert a given input
wavefront to a given output wavefront. Interferometric optical metrology of an
aspheric surface requires converting a spherical wavefront to the nonspherical
wavefront correponding to the exact shape of the surface. CGHs have a long
history, see Reference (1) and are currently used widely in optical
fabrication since many modern optical systems have one or more asphere
surfaces in them, see References (2-5). As might be guessed from the name,
CGHs are usually designed on a computer using some type of optimization
algorithm. Here we show that the CGH design for a given asphere can be derived
analytically using the grating equation. As described below, in some cases,
the analytical solution must be evaluated numerically to invert one of the
relationships since the closed form analytical solution can be very complex.
This solution also can be found analytically by iteration, depending on the
desired accuracy. 

My guess is the discussion below is common knowledge and that most everyone
already knows it and/or has worked it out for themselves. In any case here is
my derivation of the equation for the design of a CGH.

\section{Background}

The purpose of a Computer Generated Hologram (CGH), some times referred to as
a Diffractive Null Corrector, is to covert an outgoing spherical wavefront
into a wavefront which, when propagated to the surface of an asphere, is a
perfect match to the asphere surface and hence is reflected directly back onto
itself. In terms of geometrical optics this means the CGH diverts the rays in
an outgoing spherical bundle so that every ray is incident on the asphere
surface exactly along the local unit normal to the asphere surface at that
point. In this way every ray in the bundle is reflected directly back onto
itself and retraces it's way back through the CGH to become part of an
incoming spherical bundle of rays. This is illustrated in Figure 1 below.%

\begin{figure}[ptb]%
\centering
\includegraphics[
height=4.3686in,
width=5.374in
]%
{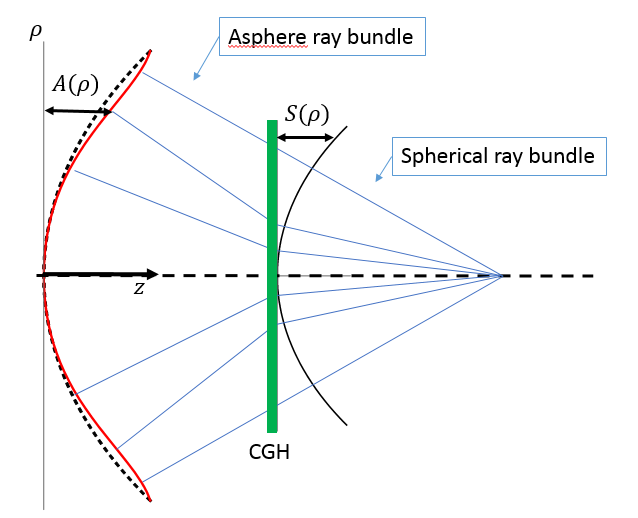}%
\caption{\textbf{Schematic of the optical layout of an asphere and a CGH, both
radially symmetric and perfectly aligned with respect to one another.
}$A\left(  \rho\right)  $\textbf{\ is the asphere surface. The dashed curved
behind it is a reference spherical surface (not the best fit spherical
surface). }$S\left(  \rho\right)  $\textbf{\ is the spherical wavefront coming
from the interferometer, and, if the CGH is designed and aligned correctly,
wavefront returning to the interferometer will be effectively perfectly
spherical for aspheres with small spherical departure.}}%
\end{figure}

Under the assumption that the asphere is radially symmetric we can use 2D
transverse $\rho$ and longitudinal $z$ coordinates. The values of $\rho$ and
$z$ are measured from the vertex of the asphere whose shape is given, as shown
in Figure 1, by%
\[
z=A\left(  \rho\right)
\]
With the CGH located a distance $L$ along the optic axis from the vertex of
the asphere, the surface of a spherical wavefront $S\left(  \rho\right)  $ of
radius $R$ with its vertex on the optic axis at the CGH, see Figure 2, is
given by
\[
z=S\left(  \rho\right)  +L=R-\sqrt{R^{2}-\rho^{2}}+L
\]

Although the CGH can be any pattern in principle, it nominally is a circular
grating with a variable period as a function of radius. The local period
$P\left(  \rho\right)  $ of the CGH grating pattern constitutes the design of
the CGH. The local period is chosen so that it meets the criteria discussed
above, i.e., it converts every ray in an outgoing spherical bundle of rays
into a ray that is incident on the apshere surface along the local unit normal
to the surface at the point of intersection. Given this criterion the design
of the CGH, i.e, the variable grating period $P\left(  \rho\right)  $ can be
determined by using the grating equation.

There are several issues that must be considered when designing a CGH. First,
although it is possible in principle to compensate for the effect of the CGH
being placed where it intersects the asphere caustic, it is much easier and
more straightforward to place the CGH outside the region of the asphere
caustic. Below we show how to calculate the caustic position in space and what
the constraint is on the placement of the CGH to avoid having it intersect the
caustic. Second, the design should make the CGH as easily manufacturable as
possible, i.e., the size of the CGH should be reasonable and the
smallest\ period or feature size in the grating pattern, should be easily
printable. Both the minimum feature size and the overall size of the CGH are
dependent on the interferometer configuration and the placement of the CGH in
the interferometer. But reasonable sizes for the CGH are on the order of a few
cm to a few 10's of cm. To avoid having to account for Maxwell electromagnetic
effects, the minimum feature size in the CGH should be greater than around 4
or 5 times the wavelength of the light used by the interferometer which
nominally will be 632.8 nm. Third, the CGH will generally be a binary
structure on glass forming either a "phase grating" or an "amplitude grating".
It follows from symmetry that binary, i.e, square wave shaped, grating
structures must have exactly the same diffraction efficiency in the $+n$ and
$-n$ diffraction orders at normal incidence. For grating periods that are
large compared to the wavelength $\lambda$ of the light, the $+n$ and $-n$
diffraction order efficiencies stay very much the same even away from normal
incidence. What this means is that even if the signal we want to detect is
just in the $+n$ order, the CGH will be generating a roughly equal amount of
$-n$ diffraction order along with other diffraction orders which we don't want
to detect. We can place an aperture near the focus of the return spherical
wave blur circle which can significantly reduce, but not completely eliminate
the effects of the unwanted orders. On the other hand a \textquotedblleft
carrier\textquotedblright\ frequency can be added to the CGH to divert the
nonzero diffraction orders off-axis or away from the origin. Which approach
should be used requires carefully evaluating the effect of the remaining
contribution from the unwanted orders on the error budget for the CGH, along
with the effect of fabrication and alignment tolerances.

Finally, one might expect the best overall CGH design involves having it, on
average, deviate the rays by the minimal possible amount but this can lead to
what might best be termed the "diffraction order swap" issue. Consider a
radially symmetric asphere and radially symmetric CGH perfectly aligned with
one another. The ray on the optic axis is undeviated and this implies the
grating period at the center of CGH should be infinite for any non-zero
diffraction order. Obviously you cannot have an infinite period on a finite
size CGH. But based on the spatial resolution of the interferometer, the
actual CGH period only needs to be the local spatial average of the "exact"
design determined using the grating equation. The spatial average reduces the
infinite period down to a finite value. The size of the spatial averaging
region should correspond, for a well designed, built and aligned
interferometer, to the interferometer spatial resolution which is the
interferometer pixel size mapped to the CGH. If the ray deviation at some
other position on the CGH surface is also zero, it also nominally requires an
infnite period at that position and most likely, considered radially, the
desired diffraction orders on opposite sides of this region will have opposite
signs, e.g., we want the +1 order on one side and $-$1 order on the other
side. As stated above an aperture placed close to the focal plane of the
spherical wavefront will automatically filter out anything that is not close
to the desired spherical wavefront.

Also note that the CGH design, which is given by the grating period as a
function of position, and the fact that it will be spatially averaged
mitigates to greater or lesser extent the difference between the intensity of
diffracted orders from so called phase and amplitude gratings. A perfect
0/$\pi$ phase grating is expected to have 0 intensity in the zeroth
diffraction order, but this is strictly true only for an infinite phase
grating. Any finite size phase grating depending on detector area, distance
from the grating, intervening optics will generally have a non-zero intensity
in the detected "zeroth diffraction order".

Figure 2 shows the return path of one of the rays starting at the surface of
the asphere. We will determine the local period of the CGH by applying the
grating equation to the return path. If the return rays precisely retrace the
outgoing path then the period determined from the outgoing rays is identical.

\textbf{NOTE}: The spherical wavefront is, by definition, radially symmetric
about the optic axis. We will assume the asphere is symmetric about the optic
axis as well and hence we can consider the CGH to be symmetric about the optic
axis. This is not necessary, but it is useful for a first analysis.

\textbf{NOTATION}:\ We will use vector notation in which $\hat{\rho}$ and
$\hat{z}$ are unit vectors in the $\rho$ and $z$ directions, respectively.

\section{Using the grating equation to determine the CGH pattern}%

\begin{figure}[ptb]%
\centering
\includegraphics[
height=4.3437in,
width=4.9796in
]%
{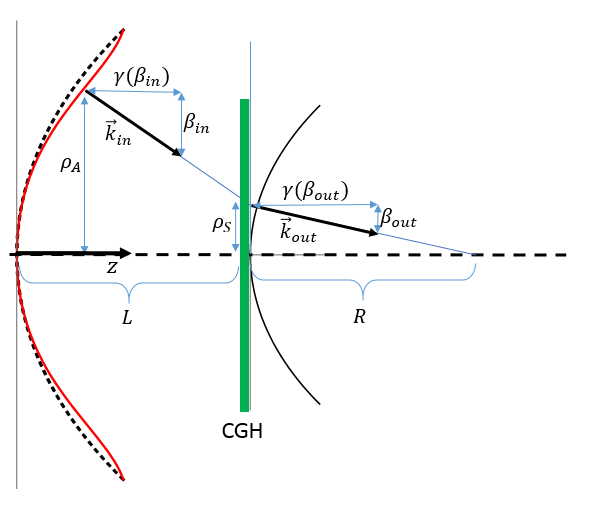}%
\caption{\textbf{Tracing one ray returning from the asphere and through the
CGH. The CGH is designed so that the return ray retraces its path back into
the interferometer. The outgoing wavefront from the interferometer is assumed
to be a perfect spherical wave with a wavefront corresponding to radius }%
$R$\textbf{\ at the position of the CGH. Since the CGH is designed so that the
rays reflected from the apshere exactly retrace their path back into the
interferometer, after trasmitting back through the CGH, the returning rays
also form a wavefront with radius }$R.$}%
\end{figure}

The local period $P$ of the CGH at position $\rho_{S},$ which we denote as
$P\left(  \rho_{S}\right)  $ must be such that $\vec{k}_{in}$ in Figure 2 is
converted to $\vec{k}_{out}$ where $\vec{k}_{out}$ corresponds to a return ray
of the spherical bundle and, at the point where it intersects the asphere
$\vec{k}_{in}$ is perpendicular to the surface of the asphere. For a return
spherical wavefront of radius $R$ as shown in the Figure 2 $\vec{k}_{out}$ is
given by
\[
\vec{k}_{out}=\beta_{out}\hat{\rho}+\gamma\left(  \beta_{out}\right)  \hat{z}
\]
with%
\[
\gamma\left(  \beta_{out}\right)  =\sqrt{k^{2}-\beta_{out}^{~2}}
\]
where $k=2\pi/\lambda.$ In terms of the propagation angle $\phi_{out}$ of
$\vec{k}_{out}$, measured with respect to the $z$ axis,
\begin{align*}
\beta_{out}  &  =-k\sin\left(  \phi_{out}\right)  =-k\frac{\rho_{S}}%
{\sqrt{R^{2}-\rho_{S}^{2}}}\\
\gamma\left(  \beta_{out}\right)   &  =k\cos\left(  \phi_{out}\right)
=\sqrt{k^{2}-\beta_{in}^{~2}}=k\sqrt{1-\left(  \frac{\rho_{S}}{R^{2}+\rho
_{S}^{2}}\right)  ^{2}}=,k\frac{R}{\sqrt{R^{2}+\rho_{S}^{2}}}%
\end{align*}
The minus sign for $\beta_{out}$ simply indicates that $\vec{k}_{out}$ is
directed toward the optic axis, i.e., in $-\hat{\rho}$ direction. The same is
true for $\beta_{in}$ for convex aspheres.

Working in terms of components of the $\vec{k}$ vectors shown in Figure 2 the
grating equation is%
\[
\beta_{out}=\beta_{in}+n\frac{2\pi}{P}
\]
where%
\begin{align*}
n  &  =\text{ grating diffraction order}\\
&  =\cdots,-1,0,1,\cdots
\end{align*}
Generally $n$ will be taken to be the first diffraction order, either positive
or negative, since this order, for most grating structures, carries the
largest intensity of diffracted light.

The unit vector normal to the asphere surface, $\hat{n}\left(  \rho
_{A}\right)  $ at position $\rho=\rho_{A}$ and pointing generically in the
$+z$ direction, can be calculated from the asphere surface shape, $A\left(
\rho\right)  ,$by%
\[
\hat{n}_{A}\left(  \rho_{A}\right)  =\left.  \frac{\vec{\partial}\left(
z-A\left(  \rho\right)  \right)  }{\left\vert \vec{\partial}\left(  z-A\left(
\rho\right)  \right)  \right\vert }\right\vert _{\rho=\rho_{A}}=\frac{\hat
{z}-\partial_{\rho_{A}}A\left(  \rho_{A}\right)  \hat{\rho}}{\sqrt{1+\left(
\partial_{\rho_{A}}A\left(  \rho_{A}\right)  \right)  ^{2}}}
\]
Taking components with respect to $\hat{z}$ and $\hat{\rho}$ we have%
\begin{align*}
\hat{n}_{A}\left(  \rho_{A}\right)  \cdot\hat{z}  &  =\frac{1}{\sqrt{1+\left(
\partial_{\rho_{A}}A\left(  \rho_{A}\right)  \right)  ^{2}}}=\cos\left(
\phi_{in}\right) \\
\hat{n}_{A}\left(  \rho_{A}\right)  \cdot\hat{\rho}  &  =-\frac{\partial
_{\rho_{A}}A\left(  \rho_{A}\right)  }{\sqrt{1+\left(  \partial_{\rho_{A}%
}A\left(  \rho_{A}\right)  \right)  ^{2}}}=-\sin\left(  \phi_{in}\right)
\end{align*}
Here $\rho,\theta,z$ are standard cylindrical coordinates with the gradient
$\vec{\partial}$ given by
\[
\vec{\partial}=\hat{\rho}\frac{\partial}{\partial\rho}+\hat{\theta}\frac
{1}{\rho}\frac{\partial}{\partial\theta}+\hat{z}\frac{\partial}{\partial z}
\]
\textbf{NOTE}: The derivatives are to be taken first and after that, set
$\rho=\rho_{A}.$

Since we are defining the setup so that the incident and reflected rays from
the asphere are everywhere along the unit normal to its surface we then have%
\begin{align*}
\vec{k}_{in}\left(  \rho_{A}\right)   &  =k\hat{n}_{A}\left(  \rho_{A}\right)
\\
&  =k\left(  \hat{n}_{A}\left(  \rho_{A}\right)  \cdot\hat{\rho}\right)
\hat{\rho}+k\left(  \hat{n}_{A}\left(  \rho_{A}\right)  \cdot\hat{z}\right)
\hat{z}\\
&  =-k\frac{\partial_{\rho_{A}}A\left(  \rho_{A}\right)  }{\sqrt{1+\left(
\partial_{\rho_{A}}A\left(  \rho_{A}\right)  \right)  ^{2}}}\hat{\rho}%
+k\frac{1}{\sqrt{1+\left(  \partial_{\rho_{A}}A\left(  \rho_{A}\right)
\right)  ^{2}}}\hat{z}\\
&  =\beta_{in}\hat{\rho}+\gamma\left(  \beta_{in}\right)  \hat{z}%
\end{align*}
The derivative with respect to the azimuthal angle $\theta$ doesn't contribute
since the radial symmetry of the asphere implies $\partial A\left(  \rho
_{A}\right)  /\partial\theta=0$.

The equation describing a ray starting at position $\rho=\rho_{A}$ and
$z=A\left(  \rho_{A}\right)  $ on the surface of the asphere and propagating a
distance $s$ along the direction of the asphere surface unit normal $\hat
{n}_{A}\left(  \rho_{A}\right)  $ to intersect the CGH at $z=L$ and $\rho
=\rho_{S}$ is, in vector notation, (see Figure 2)%
\[
\rho_{A}\hat{\rho}+A\left(  \rho_{A}\right)  \hat{z}+s\hat{n}_{A}\left(
\rho_{A}\right)  =\rho_{S}\hat{\rho}+L\hat{z}
\]
Taking the $\hat{\rho}$ and $\hat{z}$ components separately gives
\begin{align*}
\rho_{S}  &  =\rho_{A}+s\hat{n}_{A}\left(  \rho_{A}\right)  \cdot\hat{\rho}\\
L  &  =A\left(  \rho_{A}\right)  +s\hat{n}_{A}\left(  \rho_{A}\right)
\cdot\hat{z}%
\end{align*}
Solving the second equation for $s$ gives%
\[
s=\frac{L-A\left(  \rho_{A}\right)  }{\hat{n}_{A}\left(  \rho_{A}\right)
\cdot\hat{z}}=\left(  L-A\left(  \rho_{A}\right)  \right)  \sqrt{1+\left(
\partial_{\rho_{A}}A\left(  \rho_{A}\right)  \right)  ^{2}}
\]
Substituting this result into the first equation yields%
\begin{align*}
\rho_{S}\left(  \rho_{A}\right)   &  =\rho_{A}+s\left(  \rho_{A}\right)
\hat{n}_{A}\left(  \rho_{A}\right)  \cdot\hat{\rho}\\
&  =\rho_{A}+\left(  L-A\left(  \rho_{A}\right)  \right)  \frac{\hat{n}%
_{A}\left(  \rho_{A}\right)  \cdot\hat{\rho}}{\hat{n}_{A}\left(  \rho
_{A}\right)  \cdot\hat{z}}\\
&  =\rho_{A}+\left(  L-A\left(  \rho_{A}\right)  \right)  \frac{-\partial
_{\rho_{A}}A\left(  \rho_{A}\right)  /\sqrt{1+\left(  \partial_{\rho_{A}%
}A\left(  \rho_{A}\right)  \right)  ^{2}}}{1/\sqrt{1+\left(  \partial
_{\rho_{A}}A\left(  \rho_{A}\right)  \right)  ^{2}}}\\
&  =\rho_{A}-\left(  L-A\left(  \rho_{A}\right)  \right)  \partial_{\rho_{A}%
}A\left(  \rho_{A}\right)
\end{align*}
In the above we have explicitly indicated the dependence of $\phi_{in}$ on
$\rho_{A}$ since, below, we will need the inverse of this equation, i.e.,
$\rho_{A}=\rho_{A}\left(  \rho_{S}\right)  $, that is, $\rho_{A}$ in terms of
$\rho_{S.}$ This inversion can sometimes be done analytically but usually
requires a numerical solution.

Using the above relations we have that at position $\rho_{S}$ on the CGH the
period $P=P\left(  \rho_{S}\right)  $ of the CGH grating pattern must be such
that
\[
\beta_{out}=-k\frac{\rho_{S}}{\sqrt{R^{2}+\rho_{S}^{2}}}=\beta_{in}%
+n\frac{2\pi}{P\left(  \rho_{S}\right)  }=-k\frac{\partial_{\rho_{A}}A\left(
\rho_{A}\right)  }{\sqrt{1+\left(  \partial_{\rho_{A}}A\left(  \rho
_{A}\right)  \right)  ^{2}}}+n\frac{2\pi}{P\left(  \rho_{S}\right)  }
\]
The grating diffraction order $n$ has been kept unspecified for now.

Solving for $P\left(  \rho_{S}\right)  ,$ which constitutes the design of the
CGH, and substituting the formal solution for $\rho_{A}$ from above in terms
of $\rho_{S}$, i.e., $\rho_{A}=\rho_{A}\left(  \rho_{S}\right)  $ we have,
after substituting $k=2\pi/\lambda,$%
\[
P\left(  \rho_{s}\right)  =\frac{n\lambda}{\frac{\partial_{\rho_{A}}A\left(
\rho_{A}\right)  }{\sqrt{1+\left(  \partial_{\rho_{A}}A\left(  \rho
_{A}\right)  \right)  ^{2}}}-\frac{\rho_{S}}{\sqrt{R^{2}+\rho_{S}^{2}}}}
\]
As described above this result \emph{is} the CGH design.

\section{Avoiding the caustic}

The caustic is the locus of points in 3D space where adjacent rays,
propagating from the asphere surface along their respective local unit normals
to the asphere surface, intersect. For an arbitrary asphere surface, i.e., not
radially symmetric, for each point on the surface the distance to the caustic
corresponds to the smaller of the two principal radii of curvature of the
surface at that point. Since we are considering radially symmetric aspheres
the principal axes of curvature on the asphere surface are along the
tangential, $\theta$ and radial, $\rho$ directions. At position $\rho,$the
local radius of curvature in the radial or $\rho$ direction, $R_{rad}\left(
\rho\right)  $ is given by%
\[
R_{\text{rad}}\left(  \rho\right)  =\frac{\left(  1+\left(  \partial_{\rho
}A\left(  \rho\right)  \right)  ^{2}\right)  ^{3/2}}{\partial_{\rho}%
^{~2}A\left(  \rho\right)  }
\]
It follows that the longitudinal or $z$ position of the radial caustic
corresponding to radial position $\rho$ is given by
\begin{align*}
z_{C,\text{rad}}\left(  \rho\right)   &  =A\left(  \rho\right)  +R_{\text{rad}%
}\left(  \rho\right)  \hat{n}_{A}\left(  \rho\right)  \cdot\hat{z}\\
&  =A\left(  \rho\right)  +R_{\text{rad}}\left(  \rho\right)  \frac
{\vec{\partial}\left(  z-A\left(  \rho\right)  \right)  }{\left\vert
\vec{\partial}\left(  z-A\left(  \rho\right)  \right)  \right\vert }\cdot
\hat{z}\\
&  =A\left(  \rho\right)  +\frac{1+\left(  \partial_{\rho}A\left(
\rho\right)  \right)  ^{2}}{\partial_{\rho}^{~2}A\left(  \rho\right)  }%
\end{align*}
Opticians use the terminology "meridional caustic" rather than radial caustic.
The corresponding radial or $\rho$ position of the caustic corresponding to
radial position $\rho$ is given by
\begin{align*}
\rho_{C,\text{rad}}\left(  \rho\right)   &  =\rho+R_{\text{rad}}\left(
\rho\right)  \hat{n}_{A}\left(  \rho\right)  \cdot\hat{\rho}\\
&  =\rho+R_{\text{rad}}\left(  \rho\right)  \frac{\vec{\partial}\left(
z-A\left(  \rho\right)  \right)  }{\left\vert \vec{\partial}\left(  z-A\left(
\rho\right)  \right)  \right\vert }\cdot\hat{\rho}\\
&  =\rho-\frac{\partial_{\rho}A\left(  \rho\right)  \left(  1+\left(
\partial_{\rho}A\left(  \rho\right)  \right)  ^{2}\right)  }{\partial_{\rho
}^{~2}A\left(  \rho\right)  }%
\end{align*}

The local radius of curvature in the tangential direction is given by
\[
R_{\text{tan}}\left(  \rho\right)  =\frac{\rho\left(  1+\left(  \partial
_{\rho}A\left(  \rho\right)  \right)  ^{2}\right)  ^{1/2}}{\partial_{\rho
}A\left(  \rho\right)  }
\]
It follows that the longitudinal or $z$ position of the radial caustic
corresponding to radial position $\rho$ is given by
\begin{align*}
z_{C,\text{tan}}\left(  \rho\right)   &  =A\left(  \rho\right)  +R_{\text{tan}%
}\left(  \rho\right)  \hat{n}_{A}\left(  \rho\right)  \cdot\hat{z}\\
&  =A\left(  \rho\right)  +R_{\text{tan}}\left(  \rho\right)  \frac
{\vec{\partial}\left(  z-A\left(  \rho\right)  \right)  }{\left\vert
\vec{\partial}\left(  z-A\left(  \rho\right)  \right)  \right\vert }\cdot
\hat{z}\\
&  =A\left(  \rho\right)  +\frac{\rho}{\partial_{\rho}A\left(  \rho\right)  }%
\end{align*}
Opticians use the terminology "sagittal caustic" rather than tangential
caustic. The corresponding radial or $\rho$ position of the caustic
corresponding to radial position $\rho$ is given by
\begin{align*}
\rho_{C,\text{tan}}\left(  \rho\right)   &  =\rho+R_{\text{tan}}\left(
\rho\right)  \hat{n}_{A}\left(  \rho\right)  \cdot\hat{\rho}\\
&  =\rho+R_{\text{tan}}\left(  \rho\right)  \frac{\vec{\partial}\left(
z-A\left(  \rho\right)  \right)  }{\left\vert \vec{\partial}\left(  z-A\left(
\rho\right)  \right)  \right\vert }\cdot\hat{\rho}\\
&  =0
\end{align*}
The tangential caustic always occurs on the the axis of the asphere $A\left(
\rho\right)  $ but at different distances from its vertex as a function of
$\rho.$

Hence, to avoid having the CGH intersect the caustic we need $L$ to be smaller
than the minimum value of $z_{C,\text{rad}}\left(  \rho\right)  $ and
$z_{C,\text{tan}}\left(  \rho\right)  $ for all values of $\rho$ from the
center, $\rho=0,$ to the edge of the asphere, $\rho=\rho_{\text{edge}}$,%
\[
L<\min_{\rho}\left(  z_{C,\text{rad},}\left(  \rho\right)  \text{ and
}z_{C,\text{tan}}\left(  \rho\right)  \right)
\]

\textbf{NOTE}: Since the wavefront going toward the CGH from the focus at
$z=R+L$ as shown in Figures 1 and 2 ideally corresponds to a perfect spherical
wave generated by the interferometer, it follows that we want the CGH to be as
close as possible to the asphere to minimize the "non-spherical" contribution
to the returning wavefront coming from the propagation between the CGH and the
asphere and hence also minimize the fringe density. As discussed below for any
(reasonable) asphere, perfectly aligned to a perfect CGH the fringe density
generated by the asphere is zero. (This is of course the ideal case, but given
measurement uncertainties along with errors caused by unwanted diffraction
orders, etc., the ideal case can never be achieved.) But the fabrication
starts with the best fit sphere and not the perfect asphere surface and so
accurate metrology is required to determine how to figure and polish the best
fit sphere to change or move its surface to become the desired asphere
surface. The fringe density for the best fit sphere is definitely not zero and
if it is too dense then the metrology needed to proceed with the figuring and
polishing is compromised. The limitation to having the CGH as close to the
asphere as possible, i.e., having $L$ as small as possible, comes from the
maximum fabricatable size of the CGH and the sag of the asphere. Clearly from
Figures 1 and 2 the closer the CGH is to the asphere, the bigger it has to be
at least for a convex asphere.

\section{Best Fit Spheres}

There are (at least) two types of Best Fit Sphere (BFS).

The first is the opticians BFS. It is the spherical surface
\[
S_{BFS}\left(  \rho\right)  =z_{BFS}-\sqrt{R_{BFS}^{2}-\rho^{2}}
\]
for which the difference between it and the asphere surface is always
positive, i.e.,
\[
S_{BFS}\left(  \rho\right)  -A\left(  \rho\right)  \geq0\text{ : Opticians
Best Fit Sphere}
\]
for all $\rho$ from $\rho=0$ to the edge of the asphere $\rho=\rho
_{\text{edge}}.$ The reason opticians choose this definition is that they can
only remove material during figuring and polishing. So they want to start with
the closest spherical surface to the asphere suface which requires only
material removal to make the asphere.

\textbf{NOTE: }Determining the opticians best fit sphere is not a
straightforward task with a unique result for a particular asphere. It
requires the use of a nominally iterative optimization algorithm with a "merit
function", i.e., the function to be optimized, which depends in part on the
skill of the optician along with the processes used for figuring and
polishing. In the example given below where the asphere is purely parabolic, a
best fit sphere with the same radius of curvature as the parabola at it's
vertex satisfies the criteria $S_{BFS}\left(  \rho\right)  -A\left(
\rho\right)  \geq0$ for all relevant values of $\rho$ but this solution has
$S_{BFS}\left(  0\right)  -A\left(  0\right)  =0$ and so requires 0 removal of
material at the vertex which could be impossible to achieve in reality. Hence
the above condition should be replaced by something of the form $S_{BFS}%
\left(  \rho\right)  -A\left(  \rho\right)  >$ "minimum value" where this
"minimum value" depends on the skill and finesse of the optician and on the
processes he/she is using to figure and polish the surface, along with the
metrology being used to measure the surface shape during figuring and polishing.

The second so called "best fit sphere" is the Least Squares Fit (LSF)
spherical surface
\[
S_{LSF}\left(  \rho\right)  =z_{LSF}-\sqrt{R_{LSF}^{2}-\rho^{2}}
\]
which is defined by minimizing%
\[
\int_{0}^{r_{\text{edge}}}d\rho~\rho~\left(  S_{LSF}\left(  \rho\right)
-A\left(  \rho\right)  \right)  ^{2}\text{ : Least Squares "best fit sphere"}
\]
with respect to both $z_{LSF}$ and $R_{LSF}.$ The extra factor of $\rho$ in
the integrand above is because we want the best fit 2D surface and so the
weighting in the integrand increases with $\rho.$

The opticians BFS is the one that is relevant to actually fabricating the
asphere. The optician will first figure the blank to have the $S_{BFS}\left(
\rho\right)  $ surface and then use the interferometer data to figure or
polish in the asphere surface $A\left(  \rho\right)  .$ Hence the relevant
fringe density is the one obtained using the CGH designed for the surface
$A\left(  \rho\right)  $ but starting with the surface $S_{BFS}\left(
\rho\right)  .$

\section{Asphere Departure and Fringes}

The distance along a ray between the CGH and the asphere was derived above and
given by $s$. Here we relabel $s$ as $s_{A-CGH}$ just to make it clear it is
the distance from the asphere to the CGH$.$ From above we have%
\[
s\equiv s_{A-CGH}\left(  \rho_{A}\right)  =\left(  L-A\left(  \rho_{A}\right)
\right)  \sqrt{1+\left(  \partial_{\rho_{A}}A\left(  \rho_{A}\right)  \right)
^{2}}
\]
Write the asphere surface $A\left(  \rho_{A}\right)  $ as the departure
$D\left(  \rho_{A}\right)  $ from the opticians best fit sphere, i.e.,%

\[
A\left(  \rho_{A}\right)  =S_{BFS}\left(  \rho_{A}\right)  +D\left(  \rho
_{A}\right)  =z_{BFS}-\sqrt{R_{BFS}^{2}-\rho_{A}^{2}}+D\left(  \rho
_{A}\right)
\]
with
\[
D\left(  \rho_{A}\right)  \geq0\text{ for all }\rho_{A}\text{ from 0 to }%
\rho_{A,\text{edge}}
\]
where $D\left(  \rho_{A}\right)  $ is the difference between the asphere
surface and the opticians best fit sphere surface as measured in the $z$
direction. It is not the departure as measured along either the local normal
to the sphere or to the asphere.

We now have%
\[
\partial_{\rho_{A}}A\left(  \rho_{A}\right)  =\partial_{\rho_{A}}%
S_{BFS}\left(  \rho_{A}\right)  +\partial_{\rho_{A}}D\left(  \rho_{A}\right)
\equiv S_{BFS}^{\prime}\left(  \rho_{A}\right)  +D^{\prime}\left(  \rho
_{A}\right)
\]%
\[
S_{BFS}^{\prime}\left(  \rho_{A}\right)  =\frac{\rho_{A}}{\sqrt{R_{BFS}%
^{2}-\rho_{A}^{2}}}
\]

Thus%

\[
s_{A-CGH}\left(  \rho_{A}\right)  =\left(  L-S_{BFS}\left(  \rho_{A}\right)
-D\left(  \rho_{A}\right)  \right)  \sqrt{1+\left(  S_{BFS}^{\prime}\left(
\rho_{A}\right)  +D^{\prime}\left(  \rho_{A}\right)  \right)  ^{2}}
\]
and the total geometric path length along a ray that goes from the center of
the spherical wave through the CGH and to the surface of the asphere, ignoring
for now the thickness of the CGH itself%
\begin{align*}
&  s_{A-CGH}\left(  \rho_{A}\right)  +s_{CGH-R}\left(  \rho_{A}\right) \\
&  =\left(  L-S_{BFS}\left(  \rho_{A}\right)  -D\left(  \rho_{A}\right)
\right)  \sqrt{1+\left(  S_{BFS}^{\prime}\left(  \rho_{A}\right)  +D^{\prime
}\left(  \rho_{A}\right)  \right)  ^{2}}\\
&  +\sqrt{R^{2}+\left(  \rho_{A}-\left(  L-S_{BFS}\left(  \rho_{A}\right)
-D\left(  \rho_{A}\right)  \right)  \left(  S_{BFS}^{\prime}\left(  \rho
_{A}\right)  +D^{\prime}\left(  \rho_{A}\right)  \right)  \right)  ^{2}}%
\end{align*}
This result depends both on $D\left(  \rho_{A}\right)  $ and $D^{\prime
}\left(  \rho_{A}\right)  $ indicating the importance of both the spherical
departure itself, $D\left(  \rho_{A}\right)  $, and the slope of the
departure, $D^{\prime}\left(  \rho_{A}\right)  .$

\section{Accounting for the diffractive nature of the CGH}

Unfortunately the geometric path length given above is not equivalent to the
optical path length because we have not accounted for the way the CGH works.
The CGH is a periodic structure with spatially (in the case discussed here,
radially) varying periodicity. Non-zero diffraction orders can only be
understood via diffraction, they cannot be described by standard geometric ray
tracing. The diffraction orders occur due to constructive interference of the
individual wavefronts or wavelets emanating from each period of a periodic
structure. This constructive interference occurs where the OPD of the wavelets
differs by integer multiples of the wavelength. This is illustrated in Figure
3 for the first diffraction order from a transmission grating.%

\begin{figure}[ptb]%
\centering
\includegraphics[
height=2.533in,
width=4.7281in
]%
{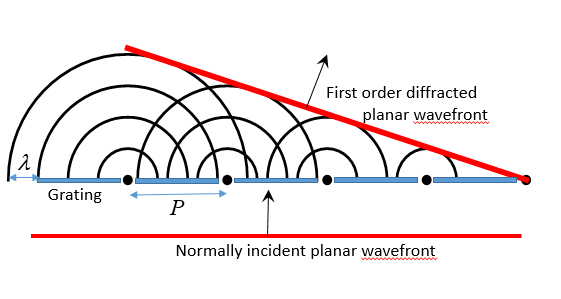}%
\caption{\textbf{The figure shows the individual wavefronts or wavelets
emanating from each aperture (shown as a "dot" of the grating structure with
period }$P.$\textbf{\ These wavelets interfere constructively to create the
first diffracted order as indicated by the tilted red line which corresponds
to the wavefront of the first diffracted order. Simple geometry shows that the
first order diffracted angle }$\theta$\textbf{\ satisfies }$\sin\left(
\theta\right)  =\lambda/P.$\textbf{\ The phases of the wavelets shown
corresponds to having a normally incident plane wave coming from below. }}%
\end{figure}

Consider the case where the first order diffracted wave is reflected back onto
itself by a plane reflecting surface as shown in Figure 4. The rediffracted
wavefront exiting the periodic structure and traveling directly downward is
planar with no phase dependence on horizontal position as indicated by the
horizontal red line representing it's wavefront. So, even though the geometric
optical path to the reflecting surface and back along the diffraction
direction varies from point to point horizontally, the diffracted wavefront
does not. Effectively, first order diffraction combined with constructive
interference removes one $\lambda$ from the geometric path for each period of
the grating. In calculating the net roundtrip OPD for the the case of a CGH
combined with an asphere we need to account for this fact to obtain the
correct fringe density.%

\begin{figure}[ptb]%
\centering
\includegraphics[
height=3.2121in,
width=4.4442in
]%
{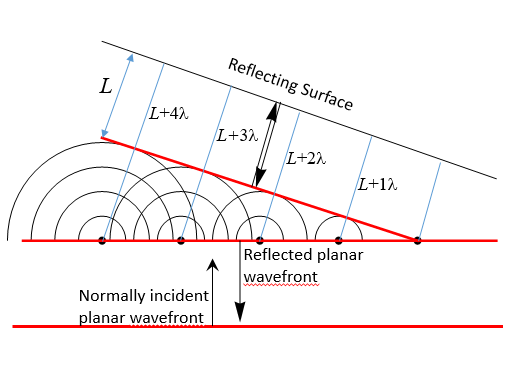}%
\caption{\textbf{The first diffracted order when reflected back onto itself
and re-diffracted through the periodic structure yields a perfectly planar or
flat wavefront with no variation in the horizontal direction propagating
directly downward even though the roundtrip optical path to the reflecting
surface varies with horizontal position. }}%
\end{figure}

It follows \ from this that the total roundtrip OPD written in terms of
$A\left(  \rho_{A}\right)  =S_{BFS}\left(  \rho_{A}\right)  +D\left(  \rho
_{A}\right)  $ is given by%
\begin{align*}
OPD\left(  \rho_{A}\right)   &  =\left(  s_{A-CGH}\left(  \rho_{A}\right)
+s_{CGH-R}\left(  \rho_{A}\right)  +C\left(  \rho_{S}\left(  \rho_{A}\right)
\right)  \right) \\
&  =2\left(  L-A\left(  \rho_{A}\right)  \right)  \sqrt{1+A^{\prime}\left(
\rho_{A}\right)  ^{2}}\\
&  +2\sqrt{R^{2}+\left(  \rho_{A}-\left(  L-A\left(  \rho_{A}\right)  \right)
A^{\prime}\left(  \rho_{A}\right)  \right)  ^{2}}\\
&  +2C\left(  \rho_{S}\left(  \rho_{A}\right)  \right)
\end{align*}
where the factors of 2 account for the roundtrip.

Consider the configuration shown in Figure 5%

\begin{figure}[ptb]%
\centering
\includegraphics[
height=3.9485in,
width=5.3333in
]%
{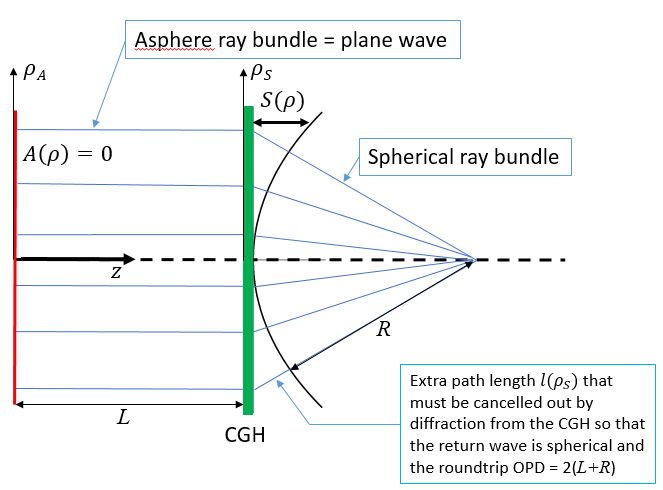}%
\caption{\textbf{In the extreme case where the "asphere" is perfectly flat it
is clear that diffraction from the CGH as discussed above must cancel out or
compensate for the extra path length labeled }$l\left(  \rho_{S}\right)
$\textbf{\ in the figure. Exactly how this works is discussed in the text. }}%
\end{figure}

As opposed to the cases shown in Figures 3 and 4 where the periodicity of the
structure, i.e., the grating, is constant, for the case shown in Figure 5 the
periodicity of the CGH structure varies with\ position $\rho_{S}.$ For first
order diffraction each period of the structure again removes one wavelength
from, or adds one wavelength to, the geometric path length depending on the
specific diffraction configuration. This compensating path length coming from
diffraction, which we label as $C\left(  \rho_{S}\right)  , $ needs to be
accounted for, and in the general case, for first order diffraction, it is
given by
\[
C\left(  \rho_{S}\right)  \equiv\int_{0}^{\rho_{S}}d\rho\frac{\lambda
}{P\left(  \rho\right)  }=\int_{0}^{\rho_{S}}d\rho\left(  \frac{\partial
_{\rho_{A}}A\left(  \rho_{A}\left(  \rho\right)  \right)  }{\sqrt{1+\left(
\partial_{\rho_{A}}A\left(  \rho_{A}\left(  \rho\right)  \right)  \right)
^{2}}}-\frac{\rho}{\sqrt{R^{2}+\rho^{2}}}\right)
\]
It follows directly from this result that by definition, $C\left(  0\right)
=0.$ Here $\rho_{S}$ is the position that counts since the CGH contains the
periodic structure and so we have used the function $\rho_{A}\left(  \rho
_{S}\right)  $ which gives the position on the asphere $\rho_{A}$ as a
function of position on the CGH $\rho_{S}.$ Of course in the integrand
$\rho_{S}$ is replaced with the dummy integration variable $\rho.$ For the
case shown in Figure 5, $\rho_{A}=\rho_{S}$ but this is obviously not the case
for general aspheres as is clear from Figures 1 and 2. For the case shown in
Figure 5, start at $\rho_{S}=0$ where, neglecting the thickness of the CGH
itself, the path length from the focus to the asphere is $R+L$. Then, in order
to calculate the correct optical or diffractive path length at position
$\rho_{S}>0$ we need to account for the\ extra path length $l\left(  \rho
_{S}\right)  $ which from simple geometry is given by
\[
l\left(  \rho_{S}\right)  =\sqrt{R^{2}+\rho_{S}^{2}}-R
\]
Since in this case $A\left(  \rho_{A}\right)  =0$, $C\left(  \rho_{S}\right)
$ reduces to
\[
-\int_{0}^{\rho_{S}}d\rho\frac{\rho}{\sqrt{R^{2}+\rho^{2}}}=-\left(
\sqrt[,]{R^{2}+\rho_{S}^{2}}-R\right)
\]
which clearly cancels the contribution $l\left(  \rho_{S}\right)  $ coming
from the geometric path length and leaves the net round trip OPD as given by
$2\left(  L+R\right)  $, which, again, neglects the CGH thickness.

\subsection{Proof that, in the ideal case, the return wavefront is perfectly
spherical}

Given the above analysis where the asphere is flat and the fact that a
perfectly spherical bundle of rays should correspond to a perfectly spherical
wavefront, since rays are defined as propagating in the direction
perpendicular to the wavefront, we come to the conclusion that for a perfectly
designed and fabricated CGH, perfeclty aligned to a given asphere, the
interferometric fringe density will be zero for the asphere, assuming of
course that it is perfect as well. (This is of course the ideal case, but
given measurement uncertainties along with errors caused by unwanted
diffraction orders, etc., the ideal case can never be achieved.) We now show
explicitly that this is the case.

The second term in $C\left(  \rho_{S}\left(  \rho_{A}\right)  \right)  $ can
be integrated directly, just as in the flat asphere case above, yielding
\[
C\left(  \rho_{S}\left(  \rho_{A}\right)  \right)  =\int_{0}^{\rho_{S}\left(
\rho_{A}\right)  }d\rho\left(  \frac{\partial_{\rho_{A}}A\left(  \rho
_{A}\left(  \rho\right)  \right)  }{\sqrt{1+\left(  \partial_{\rho_{A}%
}A\left(  \rho_{A}\left(  \rho\right)  \right)  \right)  ^{2}}}\right)
-\left(  \sqrt{R^{2}+\rho_{S}^{2}}-R\right)
\]
Noting that from previous results
\[
\rho_{A}-\left(  L-A\left(  \rho_{A}\right)  \right)  A^{\prime}\left(
\rho_{A}\right)  =\rho_{S}
\]
we see that the second term in $OPD\left(  \rho_{S}\right)  $ coming from
$s_{CGH-R}$ exactly cancels the second term in $C\left(  \rho_{S}\right)  $
above leaving just the constant $R.$ Changing integration variables in the
first term in $C\left(  \rho_{S}\left(  \rho_{A}\right)  \right)  $ from
$\rho$ to $\rho_{A}$ using the relation given above
\begin{align*}
\rho &  =\left(  \rho_{A}-\left(  L-A\left(  \rho_{A}\right)  \right)
A^{\prime}\left(  \rho_{A}\right)  \right) \\
d\rho &  =\left(  1+A^{\prime}\left(  \rho_{A}\right)  -\left(  L-A\left(
\rho_{A}\right)  \right)  A^{\prime\prime}\left(  \rho_{A}\right)  \right)
\end{align*}
where $A^{\prime\prime}\left(  \rho_{A}\right)  =\partial_{\rho_{A}}%
^{2}A\left(  \rho_{A}\right)  $ we get,
\begin{align*}
&  \int_{0}^{\rho_{S}\left(  \rho_{A}\right)  }d\rho\left(  \frac
{\partial_{\rho_{A}}A\left(  \rho_{A}\left(  \rho\right)  \right)  }%
{\sqrt{1+\left(  \partial_{\rho_{A}}A\left(  \rho_{A}\left(  \rho\right)
\right)  \right)  ^{2}}}\right) \\
&  =\int_{0}^{\rho_{A}}d\rho_{A}\left(  1+A^{\prime}\left(  \rho_{A}\right)
^{2}-\left(  L-A\left(  \rho_{A}\right)  \right)  A^{\prime\prime}\left(
\rho_{A}\right)  \right)  \frac{A^{\prime}\left(  \rho_{A}\right)  }%
{\sqrt{1+A^{\prime}\left(  \rho_{A}\right)  ^{2}}}\\
&  =\int_{0}^{\rho_{A}}d\rho_{A}~\partial_{\rho_{A}}\left(  \left(  A\left(
\rho_{A}\right)  -L\right)  \sqrt{1+A^{\prime}\left(  \rho_{A}\right)  ^{2}%
}\right) \\
&  =\left(  A\left(  \rho_{A}\right)  -L\right)  \sqrt{1+A^{\prime}\left(
\rho_{A}\right)  ^{2}}+L
\end{align*}
The first term $\left(  A\left(  \rho_{A}\right)  -L\right)  \sqrt
{1+A^{\prime}\left(  \rho_{A}\right)  ^{2}}$ exactly cancels the contribution
coming from $s_{A-CGH}$ leaving just the constant $L.$ Hence we have that, in
the ideal case,%
\[
OPD\left(  \rho_{A}\right)  =2\left(  R+L\right)  =\text{ constant}
\]
and the return wavefront going back into the interferometer is, ideally,
perfectly spherical.

Finally, as discussed above, the fabrication process starts with the opticians
best fit sphere and the fringe density for this sphere is defnitely not zero.
If its fringe density is too high then it is difficult to get stable reliable
interferometric metrology and hence difficult to determine how and where to
figure and polish the spherical surface to reshape it toward the desired
asphere surface. Also, since the outgoing rays from the CGH do not intersect
the best fit sphere at locally normal incidence the return rays do not follow
exactly the same return path which counts a re-trace error and further
complicates the analysis. Obviously in the case of fringe densities that are
beyond the resolution of the interferometer, other metrology techniques, such
as profilometry, can be used until the surface has been rehaped to be close
enough to the asphere to yield resolvable fringes. A reminder, as discussed
above, we can minimize the fringe density for the best fit sphere by placing
the CGH as close as possible to the asphere.

\section{An Example}

Let
\[
A\left(  \rho_{A}\right)  =\frac{1}{2R_{A}}\rho_{A}^{2}
\]
We will take the Best Fit Sphere (BFS), in the opticians sense, to be
\[
S_{BFS}\left(  \rho_{A}\right)  =R_{A}-\sqrt{R_{A}^{2}-\rho_{A}^{2}}
\]
where
\begin{align*}
R_{BFS}  &  =R_{A}\\
&  \text{and}\\
z_{BFS}  &  =0
\end{align*}
for this case.

As shown in Figure 6
\[
S_{BFS}\left(  \rho_{A}\right)  \geq A\left(  \rho_{A}\right)  \text{ for
}0\leq\rho_{A}\leq R_{A}
\]
From results in the previous section we have%
\begin{align*}
\rho_{S}\left(  \rho_{A}\right)   &  =\rho_{A}-\left(  L-A\left(  \rho
_{A}\right)  \right)  \partial_{\rho_{A}}A\left(  \rho_{A}\right) \\
&  =\rho_{A}-\left(  L-\frac{\rho_{A}^{2}}{2R_{A}}\right)  \frac{\rho_{A}%
}{R_{A}}%
\end{align*}
For the period of the CGH we need $\rho_{A}$ as a function of $\rho_{S}.$ Even
for this simple case the solution is rather complex%
\[
\rho_{A}\left(  \rho_{S}\right)  =\frac{R_{A}^{3}\left(  -\left(  \frac
{\sqrt{3}R_{A}^{4}\sqrt{\frac{-8L^{3}+24L^{2}R_{A}-24LR_{A}^{2}+8R_{A}%
^{3}+27R_{A}\rho_{S}^{2}}{R_{A}^{9}}}-9\rho_{S}}{R_{A}^{4}}\right)
^{2/3}\right)  -2\times3^{1/3}\left(  L-R_{A}\right)  }{3^{2/3}R_{A}\left(
\frac{\sqrt{3}R_{A}^{4}\sqrt{\frac{-8L^{3}+24L^{2}R_{A}-24LR_{A}^{2}%
+8R_{A}^{3}+27R_{A}\rho_{S}^{2}}{R_{A}^{9}}}-9\rho_{S}}{R_{A}^{4}}\right)
^{1/3}}
\]
To get a better idea of dependence of $\rho_{A}$ on $\rho_{S}$ expand the
above result in powers of $\rho_{S}$ which yields%
\[
\rho_{A}\left(  \rho_{S}\right)  =\frac{R_{A}}{R_{A}-L}\rho_{S}-\frac
{R_{A}^{2}}{2\left(  R_{A}-L\right)  ^{4}}\rho_{S}^{3}+\frac{3R_{A}^{3}%
}{4\left(  R_{A}-L\right)  ^{7}}\rho_{S}^{5}+O\left(  \rho_{S}^{7}\right)
\]
For the case considered below with $R_{A}=$ 100 mm and $L=$ 15 mm and the
maximum radius $\rho_{S}=$ $\rho_{S,\max}=$ 50 mm the first three terms in the
series shown above have the values%
\[
\rho_{A}\left(  \rho_{S,\max}=\text{ 50 mm}\right)  =\text{ 58.82}-\text{
11.97}+\text{ 7.31 }+O\left(  \rho_{S}^{7}\right)
\]
and so unfortunately for this particular case it takes many terms from the
series expansion to get a good numerical approximation.

The period, for $n=\pm1$, is thus given by
\begin{align*}
P\left(  \rho_{s}\right)   &  =\frac{\lambda}{\frac{\rho_{A}\left(  \rho
_{S}\right)  /R_{A}}{\sqrt{1+\left(  \rho_{A}\left(  \rho_{S}\right)
/R_{A}\right)  ^{2}}}-\frac{\rho_{S}}{\sqrt{R^{2}+\rho_{S}^{2}}}}\\
&  =\frac{\lambda}{\frac{\rho_{A}\left(  \rho_{S}\right)  }{\sqrt{R_{A}%
^{2}+\rho_{A}\left(  \rho_{S}\right)  ^{2}}}-\frac{\rho_{S}}{\sqrt{R^{2}%
+\rho_{S}^{2}}}}%
\end{align*}
where the overall sign is to be ignored since the period is by definition
always positive.

For the specific numerical example of an element $R_{A}=$ 100mm and maximum
radius $\rho_{S}=\rho_{S,\max}=$ 50 mm the, the maximum difference between the
two surfaces, 0.9 mm, occurs at the edge of the element $\rho_{S}=\rho
_{S,\max}$. A plot of the two surfaces is shown in Figure 6. Note that the
difference between the heights of the BFS and asphere at the edge $\rho
_{S}=\rho_{BFS}=$ 50 mm for these example values is on the order of 0.9 mm
which is very large in terms of the fringes it will produce with standard 633
nm interferometry.%

\begin{figure}[ptb]%
\centering
\includegraphics[
height=2.0921in,
width=5.8555in
]%
{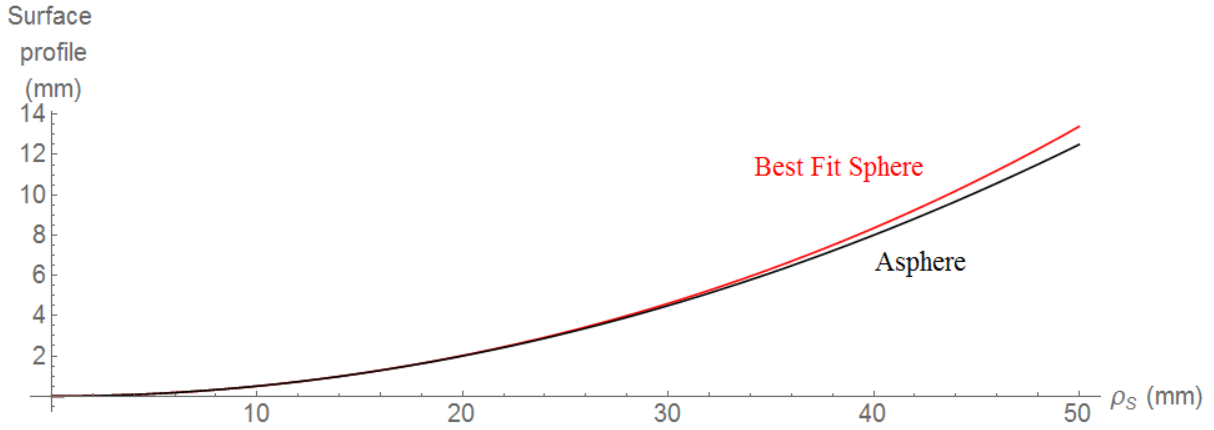}%
\caption{\textbf{Plot of the BFS surface and the asphere surface for the
specific values }$R_{A}=$\textbf{100 mm and }$\rho_{S,\max}=$\textbf{\ 50 mm.
The maximum difference of 0.9 mm occurs at the edge, }$\rho_{S,\max}.$}%
\end{figure}

The corresponding period, calculated directly from the above formula, as a
function of $\rho_{S}$ is shown in Figure 7.%

\begin{figure}[ptb]%
\centering
\includegraphics[
height=3.2495in,
width=5.8132in
]%
{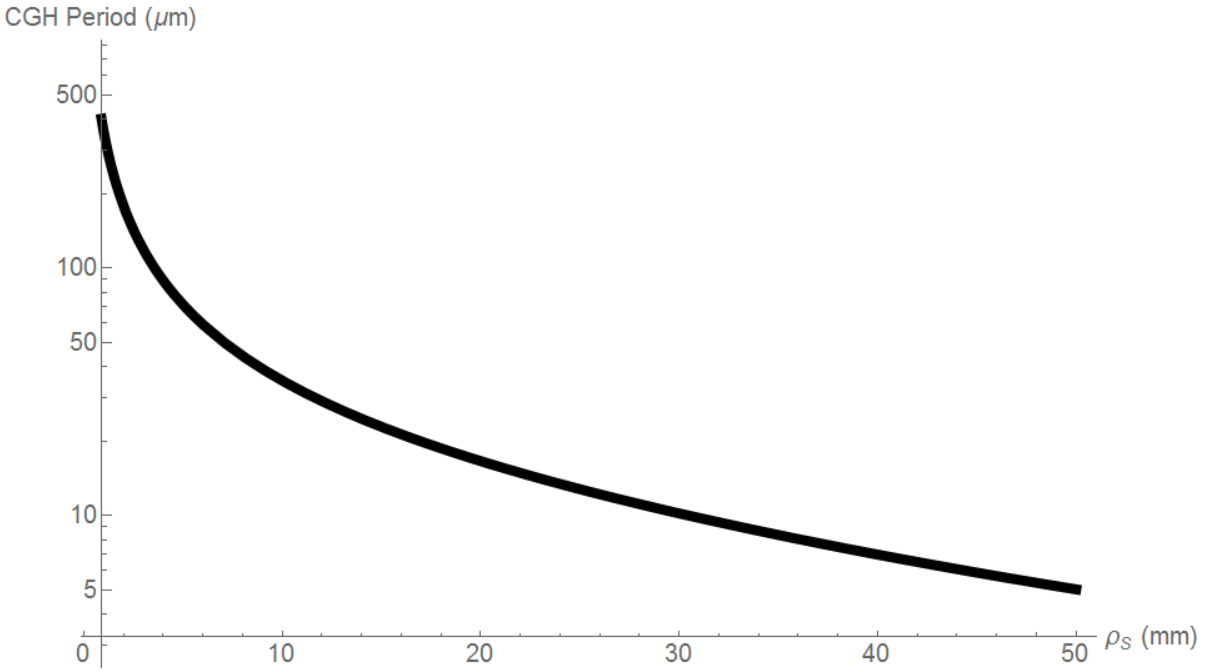}%
\caption{\textbf{The period of the CGH in microns as a function of radial
position for }$R_{A}=$\textbf{\ 100 mm, }$\rho_{S,\max}=$\textbf{\ 50 mm and
}$L=$\textbf{\ 15 mm. This is calculated directly from the formula given in
the text. NOTE: This result constitutes the design of the CGH.}}%
\end{figure}

It should be noted that strictly speaking $P\left(  \rho_{S}=0,R_{A},L\right)
=\infty$ but as shown in the graph the period drops rapidly for $\rho_{S}>0.$

The CGH is designed so that all the rays leaving the focus point of the
spherical wavefront from the interferometer at, $\rho=0$ and $z=L+R$ will
reflect directly back on themselves at the surface of the apshere and hence
will retrace their path back the focus point. For the BFS, the rays do not
reflect back on themselves because they do not intersect the BFS along its
local normal. This so called "retrace" error will be small only if the
difference between the BFS and the apshere is small. We could do an exact ray
trace to find the wavefront returning from the BFS and entering the
interferometer,\ instead we will estimate this wavefront by calculating the
distance between the BFS and the apshere along the radial direction
corresponding to the radius $R_{A}.$It is the fringe density corresponding to
this wavefront that is important for the optician at the start of figuring and
polishing and has to be interpreted correctly since he/she is using it to
determine how to figure and polish the sphere into the asphere. As mentioned
above, if this starting fringe density is beyond the resolution of the
interferometer then other metrology techniques, such as profilometry, can be
used to start the process.

The normal to the BFS surface at position $\rho=\rho_{BFS}$ on the BFS
surface, with positive component in the $z$ direction, is given by
\[
\hat{n}\left(  \rho_{BFS}\right)  =\frac{\sqrt{R_{A}^{2}-\rho_{BFS}^{2}}%
}{R_{A}}\hat{z}-\frac{\rho_{BFS}}{R_{A}}\hat{\rho}
\]
Again using vector notation the distance $l$ between the asphere and the BFS
along the normal to the BFS can be found by solving%
\[
S_{BFS}\left(  \rho_{BFS}\right)  \hat{z}+\rho_{BFS}\hat{\rho}-l~\hat
{n}\left(  \rho_{BFS}\right)  =A\left(  \rho_{A}\right)  \hat{z}+\rho_{A}%
\hat{\rho}
\]
Equating the $\hat{z}$ and $\hat{\rho}$ components separately gives%
\begin{align*}
A\left(  \rho_{A}\right)   &  =S_{BFS}\left(  \rho_{BFS}\right)  -l\frac
{\sqrt{R_{A}^{2}-\rho_{BFS}^{2}}}{R_{A}}\\
\rho_{A}  &  =\rho_{BFS}+l\frac{\rho_{BFS}}{R_{A}}%
\end{align*}
The second equation gives $l$ as%
\[
l=\frac{R_{A}}{\rho_{BFS}}\left(  \rho_{A}-\rho_{BFS}\right)
\]
Using this result to eliminate $l$ from the first equation and then solving
for $\rho_{A}$ in terms of $\rho_{BFS}$ yields%
\[
\rho_{A}\left(  \rho_{BFS}\right)  =\frac{\sqrt{R_{A}^{4}+\rho_{BFS}^{2}%
R_{A}^{2}}-\sqrt{R_{A}^{4}-\rho_{BFS}^{2}R_{A}^{2}}}{\rho_{BFS}}
\]
And so we have for $l$ in terms of $\rho_{BFS}$
\[
l\left(  \rho_{BFS}\right)  =\frac{R_{A}}{\rho_{BFS}}\left(  \frac{\sqrt
{R_{A}^{4}+\rho_{BFS}^{2}R_{A}^{2}}-\sqrt{R_{A}^{4}-\rho_{BFS}^{2}R_{A}^{2}}%
}{\rho_{BFS}}-\rho_{BFS}\right)
\]
The approximate round trip OPD this gives when converted to the number of
waves by dividing by $\lambda$ is%
\begin{align*}
OPD_{BFS,\text{approx}}\left(  \rho_{BFS}\right)   &  =2\frac{l\left(
\rho_{BFS}\right)  }{\lambda}\\
&  =\frac{2}{\lambda}\frac{R_{A}}{\rho_{BFS}}\left(  \frac{\sqrt{R_{A}%
^{4}+\rho_{BFS}^{2}R_{A}^{2}}-\sqrt{R_{A}^{4}-\rho_{BFS}^{2}R_{A}^{2}}}%
{\rho_{BFS}}-\rho_{BFS}\right)
\end{align*}
The overall factor of 2 accounts for the round-trip. The roundtrip variation
in the OPD going from the center of the BFS to the edge for 633 nm
interferometry and the values of $R_{A},$ $L$ and $\rho_{S,\max}$ given above
is on the order of 2500 waves which corresponds to roughly 5000 fringes from
center to edge which is beyond the resolution of any standard interferometer.
Increasing $R_{A}$ to 250 mm and decreasing $\rho_{S,\max}$ from 50 mm to 25
mm reduces the number of fringes to on the order of 20 from center to edge
which is more reasonable.

\section{Off-Axis case and accounting for CGH plate thickness}

Following the same procedure as above but now accounting for the plate
thickness and the interferometer focus point being off-axis as shown in Figure
8 we have%

\begin{align*}
A^{\prime}\left(  \rho_{A}\right)   &  =\partial_{\rho_{A}}A\left(  \rho
_{A}\right) \\
\vec{\beta}_{in}  &  =-k\frac{A^{\prime}\left(  \rho_{A}\right)  }%
{\sqrt{1+\left(  A^{\prime}\left(  \rho_{A}\right)  \right)  ^{2}}}\hat{\rho
}\left(  \theta\right) \\
\vec{\rho}_{A}  &  =\rho_{A}\hat{\rho}\left(  \theta\right) \\
\vec{\rho}_{S}  &  =\rho_{S}\hat{\rho}\left(  \theta\right) \\
\rho_{S}  &  =\rho_{A}-A^{\prime}\left(  \rho_{A}\right)  \left(  L-A\left(
\rho_{A}\right)  \right)  -\frac{A^{\prime}\left(  \rho_{A}\right)  }%
{n\sqrt{1+\left(  A^{\prime}\left(  \rho_{A}\right)  \right)  ^{2}}}T\\
\vec{\beta}_{out}  &  =k\frac{D\hat{\rho}\left(  0\right)  -\vec{\rho}_{s}%
}{\left\vert R\hat{z}+D\hat{\rho}\left(  0\right)  -\vec{\rho}_{s}\right\vert
}\\
&  =k\frac{D\hat{\rho}\left(  0\right)  -\rho_{S}\hat{\rho}\left(
\theta\right)  }{\sqrt{R^{2}+\left(  D\hat{\rho}\left(  0\right)  -\rho
_{S}\hat{\rho}\left(  \theta\right)  \right)  ^{2}}}\\
&  =k\frac{D\hat{\rho}\left(  0\right)  -\rho_{S}\hat{\rho}\left(
\theta\right)  }{\sqrt{R^{2}+D^{2}+\rho_{S}^{2}-2D\rho_{S}\cos\left(
\theta\right)  }}%
\end{align*}

In the off-axis case the grating equation now takes the more general form%
\[
\vec{\beta}_{out}=\vec{\beta}_{in}+\frac{2\pi}{P}\hat{p}
\]
where $\hat{p}$ is the unit vector $\left(  \hat{p}\cdot\hat{p}=1\right)  $
which everywhere points perpendicular to the local orientation of the "grating
lines" of the CGH pattern.

Rewriting the grating equation as
\[
\frac{2\pi}{P}\hat{p}=\vec{\beta}_{out}-\vec{\beta}_{in}%
\]
it follows that $\hat{p}$ points in the same direction as $\vec{\beta}%
_{out}-\vec{\beta}_{in}$ and so%
\[
\hat{p}=\frac{\vec{\beta}_{out}-\vec{\beta}_{in}}{\left\vert \vec{\beta}%
_{out}-\vec{\beta}_{in}\right\vert }%
\]
where $\left\vert \vec{\beta}_{out}-\vec{\beta}_{in}\right\vert =\sqrt{\left(
\vec{\beta}_{out}-\vec{\beta}_{in}\right)  \cdot\left(  \vec{\beta}_{out}%
-\vec{\beta}_{in}\right)  }=\sqrt{\left(  \vec{\beta}_{out}-\vec{\beta}%
_{in}\right)  ^{2}}$Taking the scalar product of the grating equation with
$\hat{p}$ gives%
\[
\frac{2\pi}{P}\hat{p}\cdot\hat{p}=\frac{2\pi}{P}=\frac{\left(  \vec{\beta
}_{out}-\vec{\beta}_{in}\right)  ^{2}}{\left\vert \vec{\beta}_{out}-\vec
{\beta}_{in}\right\vert }%
\]
Solving for $P$ then gives%
\[
P=\frac{2\pi}{\left\vert \vec{\beta}_{out}-\vec{\beta}_{in}\right\vert }%
\]
Substituting the explicit formulae for $\vec{\beta}_{out}$ and $\vec{\beta
}_{in}$ given above into the formlae for $\hat{p}$ and $P$, and expressing
$\rho_{A}$ as a function of $\rho_{S}$ then gives $\hat{p}$ and $P$ as
function of position, $\rho_{S}$ and $\theta$, on the CGH.%

\begin{figure}[ptb]%
\centering
\includegraphics[
height=3.9385in,
width=7.0809in
]%
{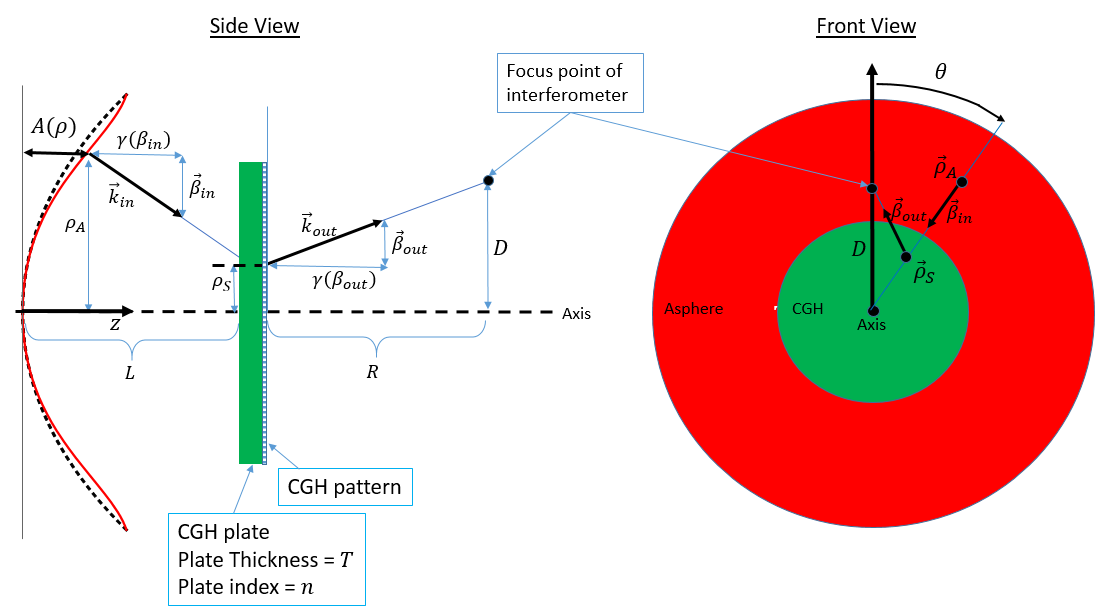}%
\caption{ }%
\end{figure}

\section{References}

\begin{enumerate}
\item J. C. Wyant and P. K. O'Neill, "Computer Generated Hologram; Null Lens
Test of Aspheric Wavefronts", Applied Optics \textbf{13}, 2762 (1974).

\item C. Pruss, S. Reichelt, H. J. Tiziani, W. Osten, "Computer-generated
holograms in interferometric testing", Optical Engineering \textbf{43(}%
11\textbf{)} (2004).

\item J-M. Asfour and A. G. Poleshchuk, "Asphere testing with a Fizeau
interferometer based ona a combined computer-generated hologram", J. Opt. Soc.
Am. A \textbf{23}(1) (2006).

\item J. C. Wyant, "Computerized Interferometric Surface Measurements",
Applied Optics \textbf{52}, 1 (2013).

\item C. Zhao, "Computer-generated hologram for optical testing: a review",
SPIE\ Proceedings, Volume 11813, 11813OK (2021).
\end{enumerate}

\end{document}